# Classicals versus Keynesians: A comprehensive table to teach 50 distinctions between two major schools of economic thought

*By* Seyyed Ali Z.N. MOOSAVIAN †

**Abstract.** Macroeconomics essentially discusses macroeconomic phenomena from the perspectives of various schools of economic thought, each of which takes different views on how macroeconomic agents make decisions and how the corresponding markets operate. Therefore, developing a clear, comprehensive understanding of how and in what ways these schools of economic thought differ is a key and a prerequisite for economics students to prosper academically and professionally in the discipline. This becomes even more crucial as economics students pursue their studies toward higher levels of education and graduate school, during which students are expected to attain higher levels of Bloom's taxonomy, including analysis, synthesis, evaluation, and creation. Teaching the distinctions and similarities of the two major schools of economic thought has never been an easy task to undertake in the classroom. Although the reason for such a hardship can be multi-fold, one reason has undoubtedly been students' lack of a holistic view on how the two mainstream economic schools of thought differ. There is strong evidence that students make smoother transition to higher levels of education after building up such groundwork, on which they can build further later on (e.g. Didia and Hasnat, 1998; Marcal and Roberts, 2001; Islam, et al., 2008; Green, et al., 2009; White, 2016). The paper starts with a visual spectrum of various schools of economic thought, and then narrows down the scope to the classical and Keynesian schools, i.e. the backbone of modern macroeconomics. Afterwards, a holistic table contrasts the two schools in terms of 50 aspects. Not only does this table help economics students enhance their comprehension, retention, and critical-thinking capability, it also benefits macroeconomic instructors to gain a holistic view and deliver such a view more easily in their classrooms. The pedagogical aspects of this approach are discussed throughout the paper with reference to the economics education literature.

## 1. Introduction

Macroeconomics explains and discusses macroeconomic phenomena from the standpoints of several schools of economic thought, each of which takes different outlooks and make different assumptions on how macroeconomic agents make decisions and how the corresponding markets operate. Therefore, building a vivid, holistic comprehension of how these schools of economic thought are distinct is a vital prerequisite for economics students to thrive academically and professionally in the discipline. Fulfilling this prerequisite becomes even more important as economics students pursue their education toward higher levels of education and graduate studies, during which students are expected to reach higher levels of Bloom's taxonomy, including analysis, synthesis, evaluation, and creation. Two major schools of economic thought are the Classical and Keynesian schools. Indeed, they have become the backbone of modern macroeconomics, and economics students can have the backbone to perform well if they comprehend the essence of these schools well.

† North Carolina State University, Department of Economics, 4102 Nelson Hall, Raleigh, NC 27695, USA.
☎. 919.515.3274 ✉. szeytoo@ncsu.edu

Teaching the distinctions and similarities of these two schools has never been an easy task to carry out in the classroom, and it becomes even more so when it comes to the graduate level. As Fernández-Villaverde from University of Pennsylvania endorses, "The first year of graduate macroeconomics is hard for aspiring economists and demands much of their instructors." Although the reason for such a hardship can be multi-fold, one reason has undoubtedly been students' lack of a holistic view on how and in what ways the two mainstream economic schools of thought differ.[2] As a result, gaining a comprehensive understanding on the major economics schools of thought will substantially help graduate economics students prosper academically and professionally.

There is strong evidence that students will make smoother transition to higher levels of education after building up such groundwork, on which they can build further later on (e.g. Didia & Hasnat, 1998; Marcal & Roberts, 2001; Islam, *et al.*, 2008; Green, *et al.*, 2009; White, 2016). Such groundwork also enables students to think critically and outside the box while they know the building blocks underneath and the assumptions underlying the arguments being made. According to Nilson (2010), "Structure is so key to how people learn" and "without structure there is no knowledge." She goes on to state that "information" is nowadays available everywhere. However, what it is not so available everywhere is organized bodies of "knowledge". She defines "information" as scattered pieces of material, whereas "knowledge" is defined by her as structured set of patterns that we have identified through observation. She also adds "Students are not stupid; they are simply novices in the discipline, who do not see the big picture of the patterns, generalizations, and abstractions that experts recognize so clearly (Arocha & Patel, 1995; DeJoneg & Ferguson-Hessler, 1996). Without such a big picture, students face another learning hurdle in addition to their other hurdles they may have." This is in fact why macroeconomics instructors must clearly and comprehensively show the structure of macroeconomic thinking in a holistic,

---

[2]. Zeytoon Nejad (2017) identifies and introduces a list of reasons why the transition from undergraduate macroeconomics to graduate macroeconomics is a hard one. These include, but not limited to, unfamiliarity with or a shortage of knowledge on economic schools of thought and their distinctions, dynamic optimization, linear algebra, economic modeling, structural modeling, log-linearization methods, stochastic processes, solving differential and difference equations, and using computer software and coding.

concise way such as the comprehensive table that the present paper puts forth.

The paper will start with a visual spectrum of various economic schools of thought. Next, the focus and scope of the paper will be narrowed down to the classical and Keynesian schools of economic thought. Afterwards, a holistic table comparing the two schools of economic though in terms of 50 aspects, including, but definitely not limited to, bases of theorization, primary methods of investigating economics phenomena, historical origin, origins of the views in terms of political-economic philosophy, time horizons of analytical vision, focuses on distinct sides of macroeconomics, main arguments, main mottos, scope of government intervention, assumptions about prices, wages, money, markets, information, expectation, interest, and unemployment, monetary policy, fiscal policy, markets, competition, micro foundation, market structure, neutrality of money, fluctuations, sources of shocks, and the like. Not only does this table benefit macroeconomic instructors to gain a holistic view and have a teaching instrument to convey such a view in their classrooms, this table can also be taken as a fine example for economics instructors on how to come up with other teaching tools similar to the pedagogical instrument introduced in this paper. Thereby, they can likewise resolve their other similar issues in delivering other economics courses as well. This table will also help economics students enhance their comprehension as well as retention when dealing with macroeconomic models. The pedagogical aspects of this approach will be discussed throughout the paper with reference to both the education literature as well as the economics education literature.

The paper is organized as follows. The next section is devoted to the introduction of important schools of economic thought as well as the new neo-classical synthesis. It also describes the evolution process of the two major schools of economic thought and the new neo-classical synthesis, and summarizes the major events occurred in the formation of these schools of economic thought. Section three presents the main discussion, which lists, tabulates, and briefly explains the primary distinctions between the two major schools. Naturally, a conclusion will follow, bringing the main points together and discussing how this approach can be utilized in other settings in order to enhance pedagogical practices.

## 2. Major schools of economic thought

A school of economic thought is a group of economists who share common ideas about economic philosophy, hold similar opinions on how the economy functions, and usually apply similar methodologies in their analyses. The main schools of economic thought that have emerged in the past few centuries include Classicals, Neo-Classicals, New Classicals, Keynesians, Neo-Keynesians, and New Keynesians, which can be classified under the two broad categories of Classicals versus Keynesians.³ Figure 1 exhibits the evolution process of the two major schools of economic thought as well as that of the new neo-classical synthesis, and summarizes a sequence of momentous events occurred in the course of the formation of these schools of economic thought.

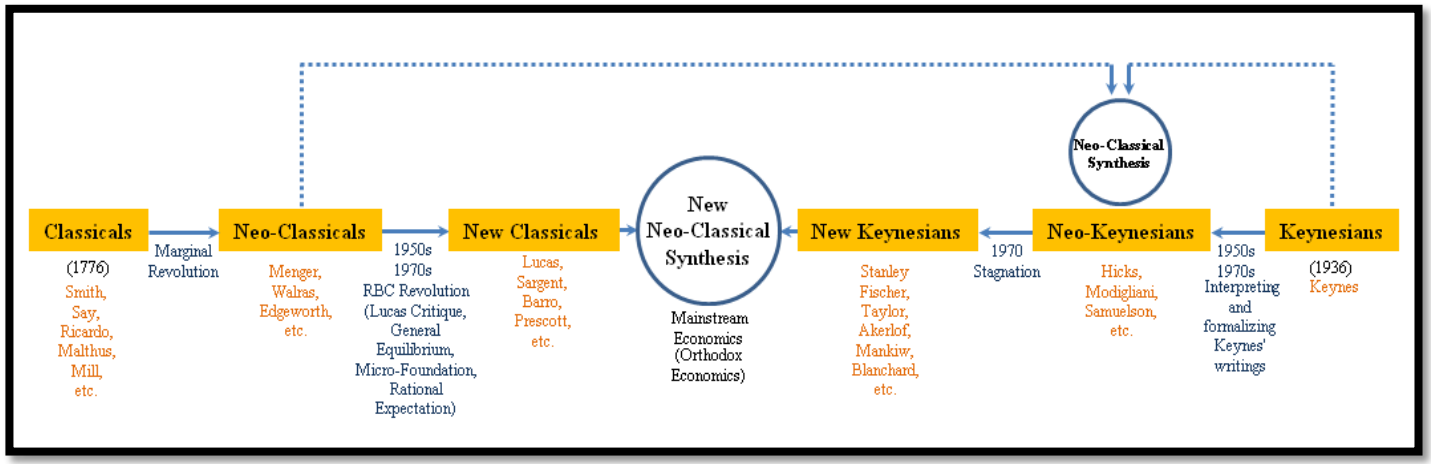

**Figure 1.** *A visual describing the evolution of the two major schools of economic thought and the new neo-classical synthesis, and summarizing the major events occurred in the formation of these schools of economic thought*
**Source:** Author's own illustration

According to Blaug (1987), classical economics (aka liberal economics) affirms that markets perform best with minimal government intervention. This school of economic thought was established in the late 18th and early 19th century by classical economists such as Adam Smith, Jean-Baptiste Say, David Ricardo, Thomas Robert Malthus, and John Stuart Mill. Adam Smith's (1776) seminal book, entitled "An Inquiry into the Nature and Causes of the Wealth of Nations," is regarded as the bible of classical economics. The main idea of his influential book is the fact that the wealth of nations, which is indeed their productive capacity, is formed on the basis of trade (free exchange of value) and not gold or other natural resources. The main difference between classical economics and modern libertarian economics is the role that they consider for the government in providing for public goods and managing common resources. Classical economists assert that markets generally regulate and adjust themselves, and often have a tendency to move towards equilibrium through an "invisible hand." They believe in the notion that private incentives are aligned with societal well-being maximization under certain competitive conditions (Blaug, 2008).

---

³. There have arisen other important schools of economic thought during the said period in the history of economics, e.g. Monetarism, whose thorough discussion is beyond the scope of the present paper. After all, many of the other schools of economic thought are somewhat close to or more or less have a general tendency toward one of the two major schools mentioned above.

Neoclassical economics is a school of economic thought that primarily focuses on the determination of goods, outputs, and income distributions in markets from the perspective of supply and demand (Campus, 1987). This determination is generally facilitated through a utility constrained maximization by individuals and profit maximization by firms given a cost function, which technically contains information on a production function, available information, and factors of production. The transition from classical economics to neoclassical economics is usually called the "marginal revolution,"[4] and has been made through the works done by economists such as William Stanley Jevons, Carl Menger, and Leon Walras.

New classical economics is a school of macroeconomic thought that conduct macroeconomic analyses fully on a neoclassical framework, and emphasizes the significance of rigorous neoclassical microeconomic foundation (i.e. micro-foundations, e.g. optimizing agents), and rational expectations, resulting in the introduction of Real Business Cycle Theory and RBC models. New classical economics is in contrast with the original Keynesian economics and Neo-Keynesian economics (to be briefly introduced in what follows), which mostly provided ad-hoc analyses, and lacked micro-foundation. New classical economics is also in contrast with new Keynesian economics (to be briefly introduced below) that uses Keynesian micro-foundations, such as nominal price rigidities and imperfect competition to create new versions of macroeconomic models, which in principle are still in line with the original Keynesian models.

Keynesian economics is a school of economic thought formed primarily based on the various existing theories about how economic output (i.e. aggregate supply) is strongly influenced by aggregate demand in the short run. Keynesian economists claim that aggregate demand can be influenced by multiple factors, and sometimes can behave very erratically, and consequently affect the levels of output, employment, and inflation (Jahan and Papageorgiou, 2014). In fact, they mean aggregate demand is not necessarily equal to the productive capacity of the economy, as argued by classical economics, and that there could be disequilibria and inefficient macroeconomic outcomes, which can be avoided or, at least, moderated by active economic policy responses, such as countercyclical monetary policy and/or countercyclical fiscal policy in order to stabilize the output level in the economy over business cycles. Keynesian economics has its original roots in John Maynard Keynes's (1936) influential book, entitled "The General Theory of Employment, Interest and Money," which founded macroeconomics as a separate branch of economics. Keynes' ideas were in contrast with those of the aggregate supply-focused classical economics preceding him.

Neo-Keynesian economics is a school of macroeconomic thought that was initially developed in the post-war period from Keynes's seminal book. This school consists of a collection of economists, such as John Hicks, Franco Modigliani, and Paul Samuelson, whose main objective was to interpret and formalize Keynes' ideas in a standard, conventional manner in economics. Subsequently, they synthesized those thoughts and ideas with the neoclassical economic models, and formed the so-called neo-classical synthesis, and created the models that shaped the fundamental ideas of neo-Keynesian economics. Keynesian economics together with Neo-Keynesians economics served as the standard macroeconomic model in the developed countries during 1940s–1970s, but they lost their popularity in the aftermath of the oil shock and stagflation of the 1970s (Fletcher, 1989).

---

[4]. Some historians of economics argue that the pace of the transition was in fact slower than the pace that the term revolution suggests (Backhouse, 2008).

In the 1970s, the appearance of a sequence of events, such as the introduction of stagflation as a newly-emerged economic phenomenon, called into question the neo-Keynesian theoretical predictions. Then, a series of new ideas (e.g. utilizing a microeconomic basis, or the so-called micro-foundation, in macroeconomic analyses) was put forth to bring novel tools to original Keynesian and Neo-Keynesian analyses, so that the new Keynesian models can explain the newly-emerged economic phenomena and events of the 1970s. The resulting school of thought was called new Keynesian economics, which, together with new Classical economics, subsequently helped the creation of the so-called "new neoclassical synthesis," which presently forms the mainstream macroeconomics (Goodfriend & King, 1997; Mankiw, 2006; Woodford, 2009).[5,6]

As mentioned before, these six schools of economic thought can be classified into the two broader categories of Classicals versus Keynesians, each of which encompasses its three respective schools. The next section will tabulate the essential distinctions among the Classical and Keynesian schools of economic thought.

### 3. Main discussion – Classicals versus Keynesians

In this section, the essential distinctions among the Classical and Keynesian schools of economic thought are discussed in greater detail in table 1. In particular, the first column (entitled "aspect") introduces the point of comparison, the second column (entitled Classicals) briefly explains the way Classicals attend to that specified aspect, and the third column (entitled Keynesians) briefly explain the way Keynesians attend to the specified aspect. The last column provides additional information and extra elaboration on related matters, if needed.

---

[5]. After the emergence of new Keynesian economics, neo-Keynesians have been sometimes called Old-Keynesians (Hayes, 2008).

[6]. The mainstream economics is sometimes referred to as "orthodox economics." In contrast, the term "heterodox economics" refers to other schools of economic thought or methodologies that are outside the "mainstream economics." "Heterodox economics" is an umbrella term used to refer to different schools, methodologies, approaches, or traditions, such as Austrian, institutional, socialist, Marxian, anarchist, evolutionary, feminist, and post-Keynesian, among others (Lawson, 2005; Lee 2008).

**Table 1.** *A table contrasting two major schools of economic thought and summarizing their distinctions*

| # | Aspect | Classicals / Neo-Classicals / New Classicals | Keynesians / Neo-Keynesians / New Keynesians | Notes |
|---|---|---|---|---|
| 1 | Primary Method of Investigating Economics Phenomena | Equilibrium Pattern | Disequilibrium Pattern | This distinction has its roots in the historical origins of their formations (prosperity vs. recession). |
| 2 | Primary Time Horizon of Analytical Vision | Long-run | Short-Run | Despite this, both also provide middle-run versions in some cases. |
| 3 | Main Argument about the Time Horizon of Analytical Vision | In the long run, prices always adjust up or down to ensure market clearing. (The concept was initially explained by Jean-Baptiste Say in 1801) | In the short run, markets do not clear. (Different markets do not clear for different reasons, e.g. goods market due to nominal price rigidity, and labor market due to efficiency wages) | - |
| 4 | Main Motto to Justify Their Time Horizons of Analytical Vision | Milton Friedman: "I don't try to forecast short-term changes in the economy. The record of economists in doing that justifies only humility." | Keynes: "In the long run, we are all dead." | McCandless & Weber (1995): If the long-run effect of monetary policy on real economic activity is truly zero, then any short-run successes in reducing downturns can only come about at the expense of reducing upturns. |
| 5 | Primary Focus of the Analysis | Economic Growth | Business Cycles and Recessions | - |
| 6 | Primary Concentration on GDP Components (Trend vs. Volatility) | General Trend of GDP, i.e. the Long-run Trend of GDP | Fluctuations of GDP (e.g. Recessions, Upturns, Prosperity, and Downturns) | Economic Value Creation vs. Economic Stabilization |
| 7 | The Side of the Economy on Which They Lay the Main Emphasis of Their Analysis | Supply Side of the Economy | Demand Side of the Economy | Moosavian (2016a and 2017) demonstrates this distinction through a comprehensive visual big picture for macroeconomics. |
| 8 | Proposed Tools to Achieve the Economic Goals Defined | Free Markets (and the Least Degree of Government Intervention) | Public Policy and Government Intervention (Monetary Policy & Fiscal Policy) | - |

**Table 1.** *A table contrasting two major schools of economic thought and summarizing their distinctions (cont.)*

| # | Aspect | Classicals / Neo-Classicals / New Classicals | Keynesians / Neo-Keynesians / New Keynesians | Notes |
|---|---|---|---|---|
| 9 | Basis of Theorization | Theorization on the basis of micro-foundation and fully optimizing agents | Traditionally, ad hoc theorization, lacking micro-foundation and lacking optimization principles, but more recently, some New Keynesian micro-foundations in models | Although the traditional Keynesian economics lacked micro-foundation, New Keynesian (NK) economics does have micro-foundation (e.g. habit persistence, menu costs, and efficiency wages) and utilizes the setup of optimizing agents. |
| 10 | The Proposed Scope of Government Intervention | The least degree of intervention | A greater extent of intervention in the economy (suggesting a more active role for the government) | - |
| 11 | Proposed Forms of Government Intervention in the Economy | Regulating property rights, and managing externalities, public goods and common resources, etc. | All tasks proposed by Neoclassicals plus monetary and fiscal policy | - |
| 12 | Origins of the View in Terms of Political-Economic Philosophy | Capitalism-Liberalism (Laissez-faire) | Capitalism-Socialism (Keynesianism) | After all, Keynesianism does not have a clear border in terms of political-economic philosophy, but definitely, it is a not a Totalitarian. |
| 13 | Their Assumptions about Prices | Changing, Fully Flexible, and Adjusting | Fixed, Rigid, and Sticky | New Keynesians point out to habit persistence, menu costs, etc. as underlying reasons and micro-foundations for price rigidities. |
| 14 | Their Assumptions about Wages | Changing, Fully Flexible, and Adjusting | Fixed, Rigid, and Sticky | New Keynesians point out to efficiency wages (multiple versions), etc. as underlying reasons and micro-foundations for wage stickiness. |
| 15 | Permanent Income Hypothesis (PIH) and Ricardian Equivalence (RI) | Holds | Does Not Hold | - |
| 16 | Quantity Theory of Money (QTM) | Holds in the long run | Does not hold in the short run | - |

**Table 1.** *A table contrasting two major schools of economic thought and summarizing their distinctions (cont.)*

| # | Aspect | Classicals / Neo-Classicals / New Classicals | Keynesians / Neo-Keynesians / New Keynesians | Notes |
|---|---|---|---|---|
| 17 | Their Assumptions about Money Neutrality | Neutrality of Money and Classical Dichotomy | Monetary-Disequilibrium Theory (Due to Bounded Rationality, Money Illusion, etc.) | Empirical evidence suggests that in the long run neutrality of money holds, but not so much in the short run. |
| 18 | Monetary Policy | No Scope (Quantity Theory of Money) | There is Some Scope. (Monetary policy has real effects in the short run.) | - |
| 19 | Fiscal Policy | There is no scope. (Because Ricardian Equivalence and Permanent Income Hypothesis hold.) | There is scope. (Because Ricardian Equivalence and Permanent Income Hypothesis do NOT hold) | NK: The assumptions of Recardian equivalence, such as capital market perfection, are not realistic. Instead, there are liquidity constraints, etc. Keynesians argue that fiscal policy can still be effective particularly in a liquidity trap, in which crowding out is trivial or absent, as interest rates do not change. |
| 20 | Aggregate Supply | Vertical | Upward-Sloping or Even Horizontal | NK: AS is upward-sloping in the middle run, and horizontal in the short run. New Classicals (NC): AS is vertical in the long run, and is a function of the existing stock of factors of production in the economy. |
| 21 | Method of Deriving IS Curve | Using Classical Cross | Using Keynesian Cross | - |
| 22 | Philips Curve | Long-Run Phillips Curve (Vertical) | Short-Run Phillips Curve (Downward-Sloping) | NK: There is a short run trade-off relationship between inflation and unemployment. NC: There is no such a trade-off in the long run. |
| 23 | Views about Central Banks and the Conduct of Monetary Policy | Monetary rules should be followed to achieve economic stability. | Discretion should be given to the Central Bank to manage the fluctuations of the economy and achieve the objectives of the dual mandate. | The monetary rule vs. discretion debate |
| 24 | Which comes first: the Supply or the Demand (A Causality Dilemma) | Say's Law: Supply creates its own demand. | Keynes' Law: Demand creates its own supply. | - |

**Table 1.** *A table contrasting two major schools of economic thought and summarizing their distinctions (cont.)*

| # | Aspect | Classicals / Neo-Classicals / New Classicals | Keynesians / Neo-Keynesians / New Keynesians | Notes |
|---|---|---|---|---|
| 25 | Main Founder | Adam Smith | John Maynard Keynes | - |
| 26 | Holy Book (The Bible) | An Inquiry into the Nature and Causes of the Wealth of Nations (1776) | The General Theory of Employment, Interest and Money (1936) | - |
| 27 | Other Economic Schools of Thought Which Have Some Close Ideas and Foundations | Classicals, Neo-Classicals, New Classicals, Monetarists, Austrians (to some degree), etc. | Keynesians, Neo-Keynesians, New Keynesians, Post Keynesians, etc. | - |
| 28 | Main Contributors to the Economic School of Thought (Other Than the Founders) | *Classicals* (Smith, Ricardo, Malthus, etc.) *Neo-Classicals* (Jevons, Menger, Walras, Edgeworth etc.) *New Classicals* (Lucas, Sargent, Barro, Prescott, etc.) | *Keynesians* (Keynes) *Neo-Keynesians* (Hicks, Modigliani, Samuelson, etc.) *New Keynesians* (Stanley Fischer, Taylor, Akerlof, Mankiw, Blanchard, etc.) | There are also post-Keynesians (Kalecki, Robinson, Kaldor, Davidson, Sraffa, Kregel, etc.), whose ideas have their roots in Keynes' book, whose discussion is beyond the scope of this table. |
| 29 | Original Universities (Academic Homeland) | University of Edinburgh, Chicago, etc. | University of Cambridge. etc. | - |
| 30 | Market-Clearing | Markets clear. | Markets do not clear. | - |
| 31 | Competition | Perfect | Imperfect/Monopolistic | - |
| 32 | Information | Perfect | Imperfect | - |
| 33 | Expectations | Rational (Forward-Looking) (Internal Model Consistency) | Keynes: Emotional NK: Rational | Rational Expectation Hypothesis implies that when information is perfect, and expectations are rationally formed, money is neutral, in the absence of money illusion. |
| 34 | Unemployment Period in the Economy | Temporary and Voluntarily | Permanent and Involuntarily | - |
| 35 | Micro Foundation | Models are always based on micro-foundations. | Historically, Keynesian economics lacked micro-foundations. More recently, NK have developed New Keynesian micro-foundation, which still uses the results from marginal revolution in the late 19th century. | Although the traditional Keynesian economics lacked micro-foundation, New Keynesian economics does have micro-foundation (e.g. habit persistence, menu costs, and efficiency wages) and uses the setup of optimizing agents. |

**Table 1.** *A table contrasting two major schools of economic thought and summarizing their distinctions (cont.)*

| # | Aspect | Classicals / Neo-Classicals / New Classicals | Keynesians / Neo-Keynesians / New Keynesians | Notes |
|---|---|---|---|---|
| 36 | Market Structure | Mostly, perfectly competitive markets, and frictionless markets | Mostly, imperfect competition, monopolistic competition, monopolies, and markets with frictions | - |
| 37 | Historical Origin and Time of Formation | Prosperity (developed in the late 18th and early 19th century) | Recession and Depression (developed during the Great Depression and in Keynes' 1936 book) | - |
| 38 | Agents' Objectives in Their Economic Models | Almost always, fully optimizers | In some cases, not fully optimizers (for a variety of reasons) | - |
| 39 | Main Creator of … | Microeconomics | Macroeconomics | - |
| 40 | Main Version of the Variable Output to Study in Their Models | Potential Output | Potential and Actual Output (Output Gap) | - |
| 41 | Neutrality or Non-Neutrality of Monetary Policy | Nearly neutral or even super-neutral (A stronger condition) | Short-run non-neutrality of monetary policy due to nominal rigidities | - |
| 42 | Competition Structure | Perfect Competition | Monopolistic Competition | - |
| 43 | Typology of the Goods Produced in Their Competitive Structures | A Homogenous Good | Differentiated Goods | - |
| 44 | Nature of Fluctuations and Volatilities | Efficiency of Business Cycles (Part of self-correcting mechanisms in the economy) | Business Cycles are Due to Market Failures (Business cycles are a reason for concern) | - |
| 45 | Primary Sources of Shocks | Technology shocks (the main source of economic fluctuations) | Various types of shocks (Monetary shocks, technology shocks, preference shocks, etc.) | - |
| 46 | Role of the School in the Development of Real Business Cycle Theory | New Classicals were the founders of RBC theory. (seminal papers: Hydland & Prescott (1982) and Prescott (1986)) | New Keynesians added Keynesians elements and features (e.g. monopolistic competition, and nominal rigidities) and developed the New Keynesian RBC model. | Gali (2015) provides a good explanation of how the RBC theory gradually formed. |

**Table 1.** *A table contrasting two major schools of economic thought and summarizing their distinctions (cont.)*

| # | Aspect | Classicals / Neo-Classicals / New Classicals | Keynesians / Neo-Keynesians / New Keynesians | Notes |
|---|---|---|---|---|
| 47 | Comparing Their RBC and DSGE Models | Basic RBC, Basic Classical Monetary Model, Money in the Utility Function (MIUF) Model, etc. | RBC model with investment cost adjustment, RBC model with habit persistence, etc. | - |
| 48 | Sources of Economic Growth | Inputs accumulation and technology progress (increased productivity) | In the short term, economic growth is caused by an increase in aggregate demand, increased government spending, etc. (given tech level and stock of inputs). | - |
| 49 | The Nature of Fluctuations in the Stock Market (Information Efficiency) | The Efficient Market Hypothesis (EMH) – Stock markets are informationally efficient. | Market Irrationality Theories (MITs) (e.g. Animal Sprits, Waves of Pessimism and Optimism, and Irrational Exuberance) | - |
| 50 | Effect of Monetary Expansion on Real Interest Rate | No effect on real interest rate (Fisher Effect holds) | Effect on real interest rate (Liquidity Effect holds) | There is empirical evidence that a positive money supply shock derives real short-term interest rate down, and real output up. (It seems a Liquidity Effect dominates in the short run, while a Fisher Effect dominates in the long run.) |
| 51 | Market Clearing | Markets clear (in the long run) | Markets do not clear (in the short run) | - |
| 52 | Views on the Relevance of Economic Policy | *Neoclassicals*: Fiscal policy causes crowding out and inefficiency, and expansionary monetary policy is irrelevant as it only causes inflation. *New Classicals* emphasize the conditions under which economic policy can be effective. | *Keynesians*: Economic policy causes real effects on the economy. Expansionary monetary policy and expansionary fiscal policy cause economic growth in the short run. Active countercyclical efforts of monetary and fiscal policy can bring about real effects and economic stabilization. Unanticipated policy has real effects. | An economic policy is countercyclical if it works against the cyclical tendencies in the economy. Expansionary monetary policy can be implemented in the form of lowering the discount rate, decreasing required reserves for banks, decreasing the interest rate paid on excess reserves, purchasing Treasury securities on the open market, etc. Expansionary fiscal policy can be implemented in the form of tax cuts, transfer payments, rebates, increased government spending, etc. |

This comprehensive table makes clear distinctions between the two major schools of economic thought by providing a point-by-point contrast of different aspects of their macroeconomic analysis. This can significantly contribute to developing a clear, comprehensive understanding of how and in what ways these schools of economic thought differ. Attaining such a holistic comprehension is a key and a prerequisite for economics students to thrive academically and professionally in the discipline, especially for economics students pursuing their studies toward higher levels of undergraduate studies as well as graduate school, during which students are typically expected to attain higher levels of Bloom's taxonomy, including analysis, synthesis, evaluation, and creation. These two major schools of economic thought are indeed the backbone of modern macroeconomics, and economics students can have the backbone to perform well if they comprehend the essence of these schools well.

Additionally, this comprehensive table makes the teaching process of these two schools much easier for economics instructors in the classroom. At the same time, this table will help economics students gain a holistic understanding on these schools of thought, which can serve as groundwork on which they can build further later on, so that they can prosper academically and professionally in the future. Such groundwork also enables students to think critically and outside the box once they know the building blocks underneath and the assumptions underlying the arguments to be made. This table will also help economics students enhance their retention when dealing with macroeconomic models. Not only does this table benefit macroeconomic instructors to gain a holistic view and have a teaching instrument to deliver such a view in their classrooms, this table can also be taken as a fine example for economics instructors on how to come up with other teaching tools similar to the pedagogical tool introduced in this paper. Thereby, they can likewise resolve their other similar issues in delivering other economics courses as well.

Naumenko & Moosvian (2016) advise that "the instructor is responsible for presenting the material in a way that would help the student navigate the knowledge in a way that is clear and digestible and eliminates room for misconceptualization." Following this advice, the present paper proposes and endorses a comprehensive table as a means of clarifying distinctions between the two major of schools of economic thought, helping the mind see and understand the patterns among the concepts and points of contrast in an organized way. The contrast table introduced in the present paper is the most comprehensive one ever put forth, as far as the author can tell. Although the paper pointed out to some theoretical aspects of the other schools of economic thought, too, the paper served to discuss only the two major ones. The other important schools will be contrasted in future papers.

There are a few points that must be taken into account when utilizing this comprehensive table. First of all, it should be noted that there are some degree of overlap between few of the aspects covered in the table, but the overlapped aspects have still been included in the table separately, thinking that it is more informative to see the distinctions clearly and in some cases in different formats in order to bring about clarity, even if it is against the principles of brevity. Secondly, the table provides a "summary" of the whole material that needs to be extensively explained and elaborated in great detail in the classroom, and it is not meant to provide "all" the details involved. Rather, it is meant to anchor the various pieces of the material which are usually taught piece by piece and separately in the classroom. As a result, it does not include all the verbal explanation and elaboration that has to be done in the classroom. Thirdly, this comprehensive table as an instructional tool can effectively serve different types of learners such as read/write leaners, visual learners, sequential learners, and global learners by providing them with texts, spatially organized material, ordered contrasts, as well as a holistic contrast, respectively.

The comprehensive contrast table proposed in this paper is an appropriate instructional tool for teaching the two major schools of economic thought for three reasons. First of all, it is a sufficiently comprehensive table, covering 50 commonly-discussed dimensions and points of comparison in macroeconomics. Secondly, the table provides complementary and explanatory notes, aiming to avoid or clear up confusions about the distinctions explained. Further, the table comes with a supporting visual describing the evolution of the two major schools of economic thought, and summarizing the major events occurred in the formation of these schools of economic thought, which gives some underlying structure to the schools covered in the table.

Quite often, macroeconomics textbooks introduce these distinctions in separate chapters and sections, if they do so explicitly. Doing so obscures the understanding of the linkages among these connected concepts. Failure to provide students with a "big picture" of the distinctions underlying the arguments, reasoning, and analyses may cause the material to become complicated when the students are expected to know the whole picture, which in turn can contribute to weakening students' analytical ability. Providing such a comprehensive table can have multiple advantages not only for students but also for instructors. Some, but not all, of the advantages are as the following: creating a strong mental framework of the distinctions between the schools, preventing students from getting lost among various schools of economic thought, increasing comprehension level, enabling students to conduct holistic analyses by understanding all the aspects involved in the analysis, allowing instructors to take a multimodal approach to their teaching of macroeconomics, providing a tool to discuss pluralistic ideas in teaching macroeconomics, etc.

In order for a comprehensive contrast table to be considered well-designed, there are multiple handy tips that should be followed. Moosavian (2016a) provides somewhat similar tips for designing visual "big pictures," some of which apply to comprehensive contrast tables as well. Here is a short list of applicable, useful tips on designing a comprehensive contrast table: Remain consistent with the terminology used. Avoid going into much detail.

Design on the basis of the learning objectives of the courses, if possible. Put necessary notes that help avoid confusions. Take advantage of color-coding if possible or needed. Sometimes, even apparently minor things matter, so consider them in your design. Tidiness matters a lot. There is always a trade-off between holding simplicity and elaborating complexity, so try to find the optimal combination of the two. Ask for help for the parts you have no idea on how to get them done. These are indeed the tips that one can take advantage of in practice when designing his or her own comprehensive contrast table. The next section of the paper draws a conclusion of the whole discussion offered in this paper.

## 4. Conclusion

Economics is a discipline that involves a wide variety of schools of thought, each of which can differ from the others in terms of numerous aspects. This variety and numerousness can be a source of confusion for many economics students. This paper proposes a holistic contrast table that can serve as a complementary tool to eliminate this potential reason for concern in the teaching of macroeconomics. It is suggested that a typical contrast table can help students avoid confusions and mental chaos caused by the plurality of the schools and the variety of the distinctions. It is shown that this important task can be undertaken through structuring an organized contrast table, which clearly explains all the existing distinctions, helping students clear up the fuzziness and confusions about the existing distinctions. In particular, the primary aim of this paper is to facilitate the teaching and learning of the two major schools of economic thought, i.e. Classicals and Keynesians, through the use of a holistic contrast table, which can be used as a complimentary resource in macroeconomics classes, aiming at helping students to get the big picture of the distinctions at once in a single table. The potential that such a table has to improve teaching practices and enhance learning when complementing traditional context is discussed under the context of contemporary teaching and learning literature. Embedded throughout the paper are suggestions and handy tips for how instructors should design and apply such a tool in their teaching.

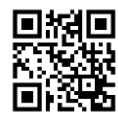

S.A.Z.N. Moosavian, JEST, 9(2), 2022, p.63-79.